\documentclass[superscriptaddress,showpacs,twocolumn,aps,pre,10pt]{revtex4-1}
\pdfoutput=1
\usepackage{natbib}
\usepackage{graphicx,dcolumn,bm,bbm,amsmath,amssymb,color}
\usepackage{caption}
\usepackage{subcaption}
\usepackage{xfrac}
\usepackage{breqn}
\usepackage{color,soul}

\newcommand{\eeq}{ \end{equation} }
\newcommand{\beq}{ \begin{equation} }
\newcommand{\bea}{\begin{eqnarray}}
\newcommand{\eea}{\end{eqnarray}}

\makeatletter
\let\cat@comma@active\@empty
\makeatother
 
\begin{document}
\title{Two dimensional self-assembly of inverse patchy colloids}
 
\author{Remya Ann Mathews K}
\author{Ethayaraja Mani*}
\affiliation{Polymer Engineering and Colloid Science Lab, Department of Chemical Engineering, Indian Institute of Technology Madras, Chennai - 600036, India}
 \email{ethaya@iitm.ac.in}
\date{\today}

\begin{abstract}
We report on the self-assembly of inverse patchy colloids (IPC) using Monte Carlo simulations in two-dimensions. The IPC model considered in this work corresponds to either bipolar colloids or colloids decorated with complementary DNA on their surfaces, where only patch and non-patch parts attract. The patch coverage is found to be a dominant factor in deciding equilibrium self-assembled structures. In particular, both regular square and triangular crystals are found to be stable at 0.5 patch coverage. Upon decreasing the patch coverage to 0.33, the regular square crystal is destablized; instead rhombic and triangular crystals are found to be stable. At low patch coverages such as 0.22 and 0.12, only triangular crystal is stabilized at high density. Particles of all the patch coverages show kinetically stable cluster phases of different shape and size, and the average cluster sizes are found to strongly depend on the patch coverage and particle density. State-diagrams showing all the stable phases for each patch coverage are presented. Ordered phases are characterized by bond order parameters $\psi_4$, $\psi_6$ and radial distribution function. The effect on patch coverage on polarization of the stable structures are also studied. The study demonstrates that inverse patchy colloids are potential candidates to form various ordered two-dimensional structures by tuning the size of the patch.
\end{abstract}
 
\maketitle
\section{Introduction}
Self-assembly of colloidal particles into various complex structures has attracted much attention in recent times. Great strides in the synthesis of anisotropic colloids in terms of shape and surface chemistry has broadened the avenues for bottom-up assembly \cite{glotzer2007anisotropy,pawar2010fabrication,sabapathy2015synthesis,sabapathy2016synthesis}. In tandem, progress has been made in engineering inter-particle interactions via selective hydrophobic coating \cite{romano2011two,sabapathy2015synthesis,hong2008clusters,li2013self}, DNA grafting \cite{nykypanchuk2008dna,park2008dna,xing2011dna,feng2013dna}, and by creating asymmetric charge distribution \cite{hong2008clusters,bianchi2014tunable}, to mention a few. These patches on the colloidal particles can be tuned to be attractive or repulsive. The main feature of these particles with orientation-dependent interactions, like the molecular systems, is that they can be used to self-assemble into complex ordered structures. A much studied model system in this context is Janus particles, with half of its surface coated with an attractive patch \cite{sciortino2009phase,sciortino2010numerical,vissers2013predicting}. 
\\

The number, size and disposition of patches are shown to affect the phase behavior of these particles. For instance, the vapor-liquid equilibrium shifts to lower temperatures as the patch size decreases \cite{kern2003fluid}. The Kern-Frenkel model designed for orientation dependent interactions of hard spheres with attractive patches, has opened up a new horizon to model patchy particles \cite{kern2003fluid}. Depending upon the number of patches, which determine the number of bonds per particle, various phases such as  polydisperse chains, bonded planes, sheets, ordered triangular and square crystals are stabilized \cite{giacometti2010effects,mani2012sheet}. The directional nature of these patches are also useful in modelling biological phenomena, for instance protein crystallization \cite{liu2009self}. Like surfactants, Janus particles are shown to stabilize clusters mimicking the structure of micelles and vesicles under certain conditions \cite{sciortino2010numerical}. A soft Janus model has been reported to stabilize superstructure like micelles, wormlike strings, helices, and bilayers depending upon the strength of patchy attraction and patch size \cite{li2012model}. A two-patch model with different patch coverages that allow two bonds per patch has been shown to stabilize Kagome crytsal \cite{romano2011two}.\\

A simple patchy model system that can be experimentally realized, yet not much studies have been reported is the inverse patchy colloid (IPC) model. Unlike the Janus colloids where patch-patch attraction takes place, in the IPC, patch and non-patch surfaces attract.  \citeauthor{hong2006clusters} conducted self assembly experiments with 'zwitterionic colloids' wherein each particle carries positive charges on one-half and negative charges on the other half. Experimentally observed clusters consisting upto 13 particles were reproduced in simulations using a patchy model \cite{hong2006clusters}. Another IPC model that has been recently studied consists of a charged particle with a positively charged equatorial region and two equal negatively charged patches at the poles \cite{bianchi2006phase,bianchi2014tunable,van2015simple}. The coarse-grained model of this sytem is compared with and mapped onto an analytical pair potential derived based on the Debye-Huckel approach \cite{bianchi2011inverse}. These IPCs self-assemble into parallel crystalline monolayers in bulk and into microcrystalline gels under confinement \cite{noya2014phase,bianchi2013self}. Analytical studies of equilibrium properties like internal energy and pair distribution function have also been carried out on IPCs with variable number of patches \cite{kalyuzhnyi2015inverse}. IPC models have recently been experimentally realized in DNA coated colloids, wherein a particle is selectively coated with single-stranded DNA and its complementary DNA such that patch and non-patch attraction can be induced, while patch-patch orientation is repelled due to steric effects \cite{feng2013dna}.\\

Recently \citeauthor{dempster2016aggregation}, reported a simulation study on the aggregation of heterogeneously charged  colloids in three-dimensions. They reported the formation of polarized close-packed crystal at Janus case, while wormlike glass, size-limited isotropic clusters and cross-linked gels are formed with successive decrease in patch coverage. The polarisation of structures at high patch coverage shows a first order decrease at larger values of polydispersity in terms of patch boundary and size. They also discuss the liquid-vapor coexistence at high patch coverage while addition of isotropic colloids is found to favour an otherwise kinetically inaccesible phase-segregation at low patch coverage \cite{dempster2016aggregation}. 

We have recently reported experimental methods to synthesize bipolar particles with variable patch coverage \cite{C7CP00680B,sabapathy2016synthesis,sabapathy2015visualization}. The bipolar particles are synthesized by coating a negatively charged patch on an otherwise positively charged particle. By choosing various interfaces such as air-water, oil-water etc., the width of the patch is altered due to the change in the three phase contact angle. Inspired from our recent experiments, we report in this paper, a Monte Carlo simulation study on the two-dimensional self-assembly of IPC. In particular we investigate the effect of patch coverage, interaction strength and particle density on the phase behavior of IPC.  

\section{Model \& Simulation}

\begin{figure}[h]
     		\centering
		\includegraphics[height=3.5cm]{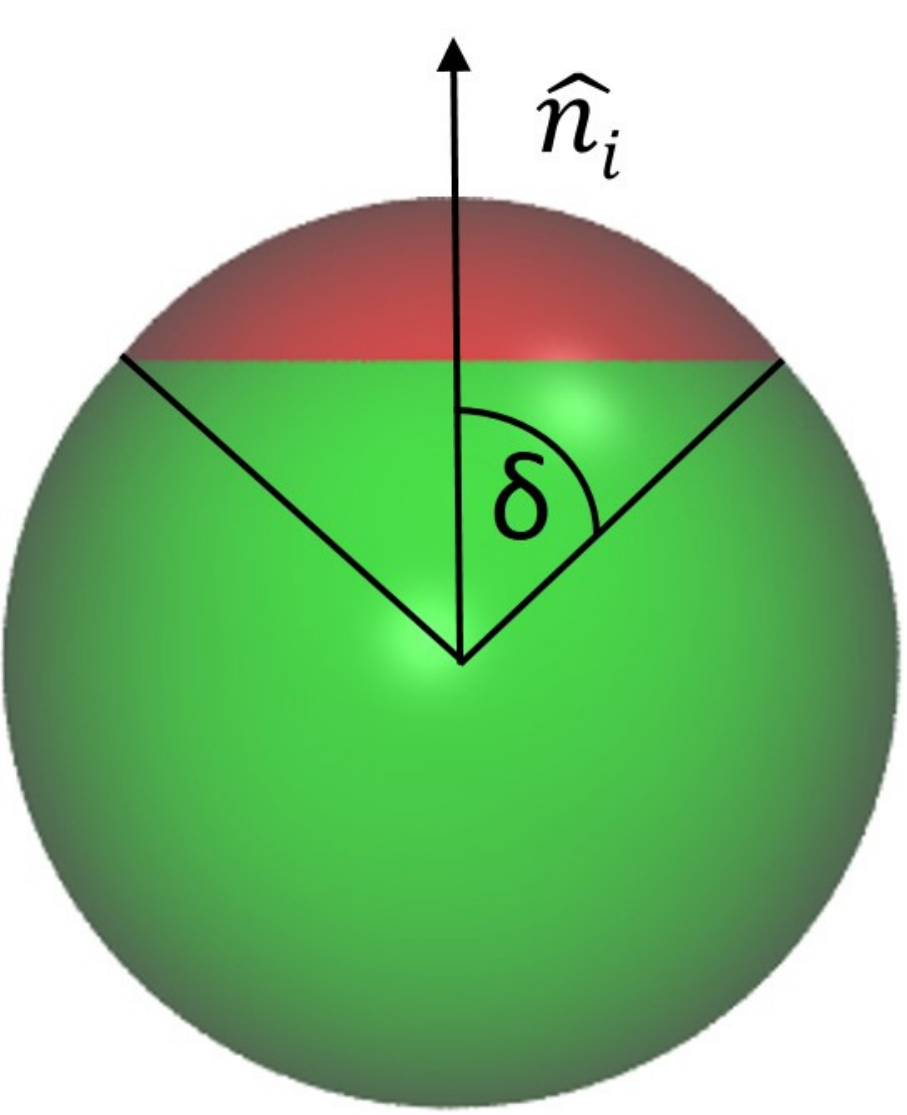}
		\captionsetup{justification=raggedright}
     		\caption{Schematic of a inverse patchy colloid. The orientation is denoted by the arrow. $\delta$ is the patch angle. (color online)}
     		\label{ipcparticle}
\end{figure}

The IPC is modelled as a hard disk of diameter $\sigma$ with a patch characterized by an opening angle of 2$\delta$. Here $\delta$ is called the patch angle, which determines the width of the patch ($\chi$) on the particle as shown in Fig.~\ref{ipcparticle}.
\\ We assume a positively charged particle with a negatively charged patch, hence it behaves as a bipolar particle. The model can also be used for complementary DNA-coated IPC. Both of these scenarios correspond to patch to non-patch attractive interactions. A unit vector $\mathbf{\hat n}_{i}$ is drawn from the center of the particle through the center of the patch to represent the orientation of the particle. The arc length covered by the patch or the patch coverage ($\chi$) can be expressed as
\begin{equation}
\chi = \delta/\pi
\end{equation} 

\begin{figure}[h]
\includegraphics[height=10cm]{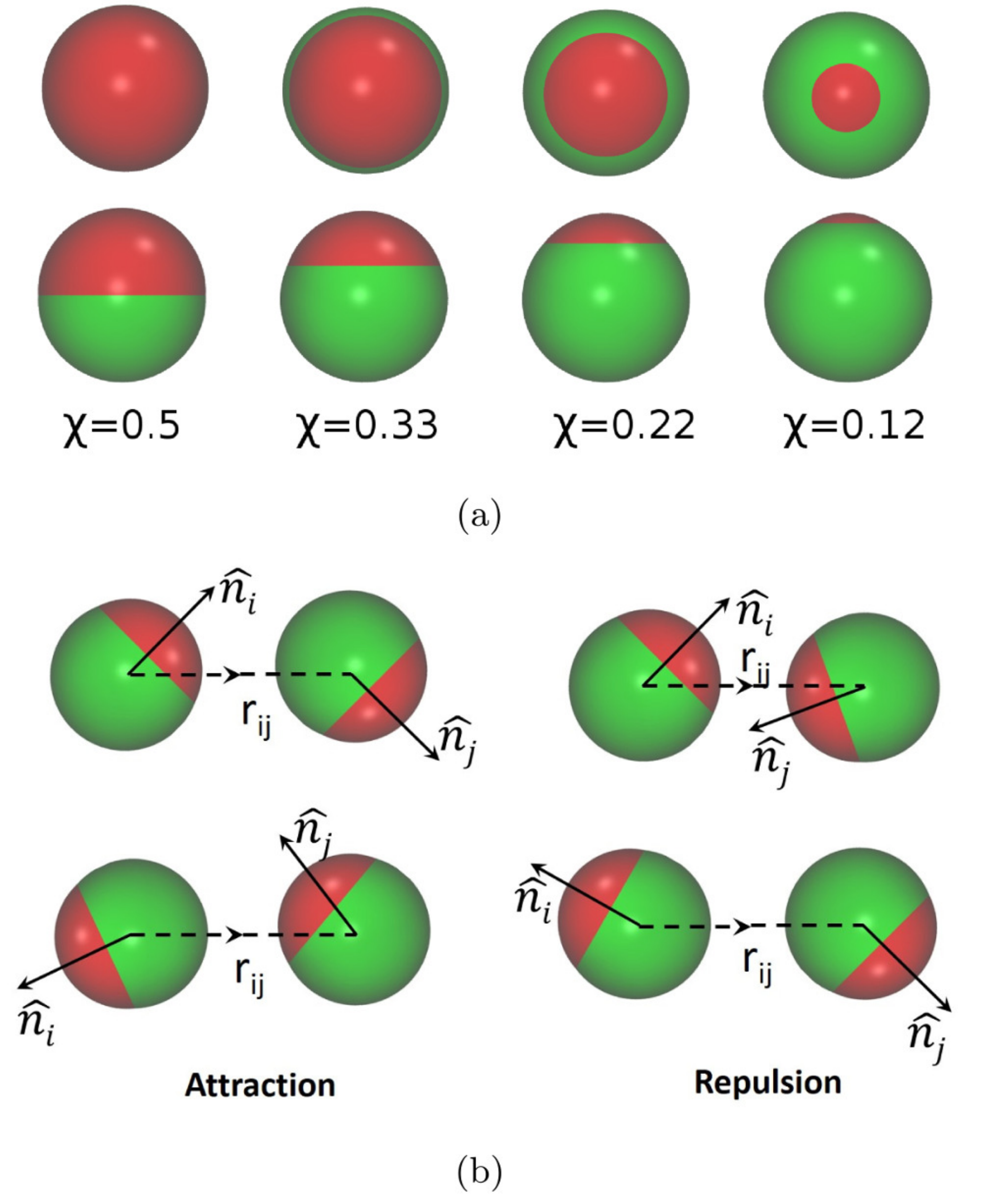}
\captionsetup{justification=raggedright}
\caption{(a) Particles with different patch coverages. Top view(top row), side view (bottom row). Green color shows positive part of the particle and red color shows the negative patch on it. (b) Schematic of the interaction between two patchy particles. r$_{ij}$ is the distance between the two particles i and j. $\mathbf{\hat n}_i$ and $\mathbf{\hat n}_j$ are the unit vectors that define the orientation of the paticles i and j respectively. Different orientations leading to attractive and repulsive interactions are depicted. (color online)}
\label{patch}
\end{figure}

The effect of patch coverage on the phase behavior of IPC is studied by considering four different patch coverages as shown in Fig.~\ref{patch}a. The distance between two particles and the orientation of their patches decide whether they would attract or repel each other. A interaction range of 50$\%$ of the particle diameter ($\lambda = 1.5\sigma$) is chosen. The model incorporates a square well interaction $U^{sw}(r_{ij})$ for the strength, and an orientation dependent term $ \Omega (\mathbf{\hat n}_{i} ,\mathbf{\hat n}_{j}, \mathbf{\hat r}_{ij} )$. When the distance between the center of two particles comes in the interaction range of $\sigma$ to $ \lambda\sigma$, the particles experience pair interactions. Let $\theta_i$ is the angle between the patch vector $\mathbf{\hat n}_{i}$ of particle $i$ and the vector joining the centers of particles $i$ and $j$ ($\mathbf{\hat r}_{ij}$), and let  $\theta_j$ is the angle between the patch vector $\mathbf{\hat n}_{j}$ of particle $j$ and the vector joining the centers of particles $j$ and $i$ ($\mathbf{\hat r}_{ji}$). If both $\theta_i$ and $\theta_j$ are either greater or less than $\delta$, then the particles are repulsive, and attractive otherwise. In other words, like sides of two particles would repel each other while unlike sides would attract, within the range of interaction as shown in Fig.~\ref{patch}b. The model is written mathematically as shown in Eq.~(\ref{equation1}).

\begin{equation}
	U_{ij} = U^{sw}(r_{ij})  . \Omega (\mathbf{\hat n}_{i} ,\mathbf{\hat n}_{j}, \mathbf{\hat r}_{ij} )
\label{equation1}
\end{equation}
where,\\

\begin{equation}
	U^{sw}(r_{ij}) = 
	\begin{cases}
		\infty & \text{if $\sigma \geq  r_{ij}  $} \\ 
		\epsilon & \text{if $\sigma < r_{ij} \leq \lambda\sigma $} \\  
		0 & \text{if $\lambda\sigma < r_{ij} $} \\
	\end{cases}
\end{equation}

\begin{dmath}
	\Omega (\mathbf{\hat n}_{i} ,\mathbf{\hat n}_{j}, \mathbf{\hat r}_{ij} ) =
	\begin{cases}
		-1   \begin{cases}
						\text{$\mathbf{\hat r}_{ij}.\mathbf{\hat n}_{i} \leq cos \delta$ and $\mathbf{\hat r}_{ji}.\mathbf{\hat n}_{j} \geq cos \delta$} \\
						\text{$\mathbf{\hat r}_{ij}.\mathbf{\hat n}_{i} \geq cos \delta$ and $\mathbf{\hat r}_{ji}.\mathbf{\hat n}_{j} \leq cos \delta$}\\
					\end{cases}\\
		+1    \begin{cases}
						\text{$\mathbf{\hat r}_{ij}.\mathbf{\hat n}_{i} \leq cos \delta$ and $\mathbf{\hat r}_{ji}.\mathbf{\hat n}_{j} \leq cos \delta$}\\
						\text{$\mathbf{\hat r}_{ij}.\mathbf{\hat n}_{i} \geq cos \delta$ and $\mathbf{\hat r}_{ji}.\mathbf{\hat n}_{j} \geq cos \delta$}\\
					\end{cases}
	\end{cases}
\end{dmath}

This model considers a general, single patch IPC, wherein the interactions can be either electrostatic or DNA-mediated. The model being dependent on the patch coverage and orientation, it becomes important to understand the effective interactions (orientationally averaged) between the particles. We calculate second virial coefficient for the above model for this purpose. We follow the derivation of $B_2$ calculation for a patchy particle in three-dimensions \citep{kern2003fluid}, and adopt it for our model in two-dimensions. The expression for  $B_2$ is given as:
\begin{dmath}
	\frac{B_2^{IPC}}{B_2^{HS}} = 1-\Big[\lambda^2-1\Big]\Bigg[\Big(\chi^2+(1-\chi)^2\Big)(e^{-\beta\epsilon}-1)+2\chi(1-\chi)(e^{\beta\epsilon}-1)\Bigg]
\end{dmath}

where, $\beta = 1/k_BT$, $B_2^{IPC}$ is the second virial coefficient for an IPC and $B_2^{HS} = \pi\sigma^2/2$ represents the second virial coefficient of a hard-disk of diameter $\sigma$. We find that $B_2^{IPC}$ is negative for $\epsilon > 2.1$. (The crossover points are: for $\chi$ = 0.5 $\epsilon =  1.15$, $\chi$ = 0.33 $\epsilon =  1.3$, $\chi$ = 0.22 $\epsilon =  1.55 $, $\chi$ = 0.12 $\epsilon =   2.1$). In our simulations, we maintain $\epsilon$ values above the zero cross over values for the respective patch coverage.  \\

We perform Monte Carlo simulations in two-dimensions using the NPT ensemble. We trace the state diagram of a system of N = 500 particles by varying the particle number density, $\rho_{A}$ ($\sigma^2$ units) and the strength of  interaction, $\epsilon$ ($k_{B}$T units).  We set $k_BT = 1$. An increase in $\epsilon$ can be considered as a corresponding decrease in temperature. For lower densities (upto 0.6), the system is started with a random initial configuration. In the case of higher densities ($>$ 0.7), we start with a square lattice which is then melted at a low attraction strength to produce an equivalent random initial configuration. The simulation is done at different initial configurations and equilibrated to maintain consistency of the results.  Each system is allowed to equilibrate upto a minimum of $3\times10^{8}$ MC cycles followed by another $2\times10^{6}$ MC cycles during which we study the behaviour of the system. Each MC cycle consists of N particle moves and a volume change move. Each particle move consists of a translation and rotation following the algorithm given by \citeauthor{frenkel1996understanding} \cite{frenkel1996understanding}. The acceptance ratio during equilibration is maintained between 0.4-0.6. The particle is allowed to rotate within a range of $\pm10\%$ of its direction in the x-y plane. The system is considered to be in equilibrium when the energy per particle and density attain equilibrium values and fluctuate around mean values. Simulations in NVT ensemble were performed to confirm the stability of phases observed at a particular value of $\epsilon$ and the density.

Structural properties like radial distribution function, and the collective orientations or polarization of particles in the ordered phases found in each system are discussed. The variation of average cluster size with respect to patch coverage is also reported to justify the change in nature of the clusters. A crystal order parameter $\psi_n$ can be used to measure the relative order within a given crystal. Mathematically it is expressed as \cite{steinhardt1983bond,bianchi2014tunable},
\begin{equation}
	\psi_{n} = \Bigg \langle\frac{1}{N}\sum\limits_{i=1}^N\Biggl| \Bigg \langle\frac{1}{N_n}\sum\limits_{j=1}^{N_n} e^{in\theta_{ij}} \Bigg \rangle \Biggr| \Bigg \rangle 
\label{chivalue}
\end{equation}

where N represents the total number of particles, and $N_n$ gives the nearest neighbors of the i$^{th}$ particle in the interactive range. $\theta_{ij}$ represents the angle between the vector joining the centers of particles $i$ and $j$ and an arbitrary direction chosen.

\section{Results and Discussion}

From NPT and NVT Monte Carlo simulations, we construct the state diagram for various patch coverages. The phases stabilized at different patch coverages depend crucially on the patch coverage, strength of interaction and particle density. Subsequently, the structural and orientational ordering of the ordered crystal phases for each patch coverage are discussed in detail.

\subsection{State diagram}
\begin{figure*}[t!]
\centering
\includegraphics[height=13cm]{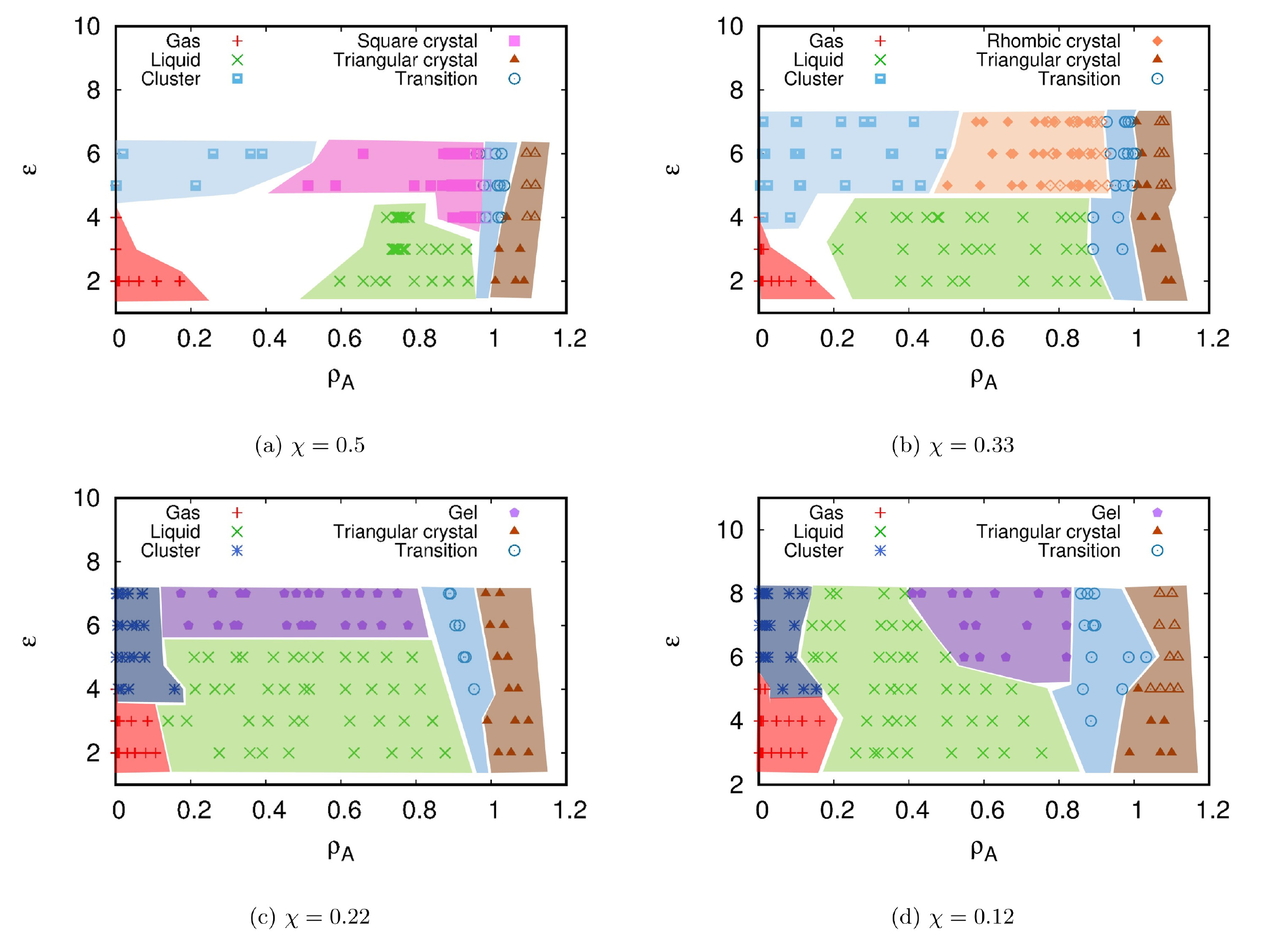}
   \captionsetup{justification=raggedright}
   \caption{State diagram for the system at different patch coverages showing gas, liquid, cluster and crystal phases. (color online)}
   \label{statediagram}
\end{figure*}

\begin{figure*}[t!]
\centering
\includegraphics[height=17cm]{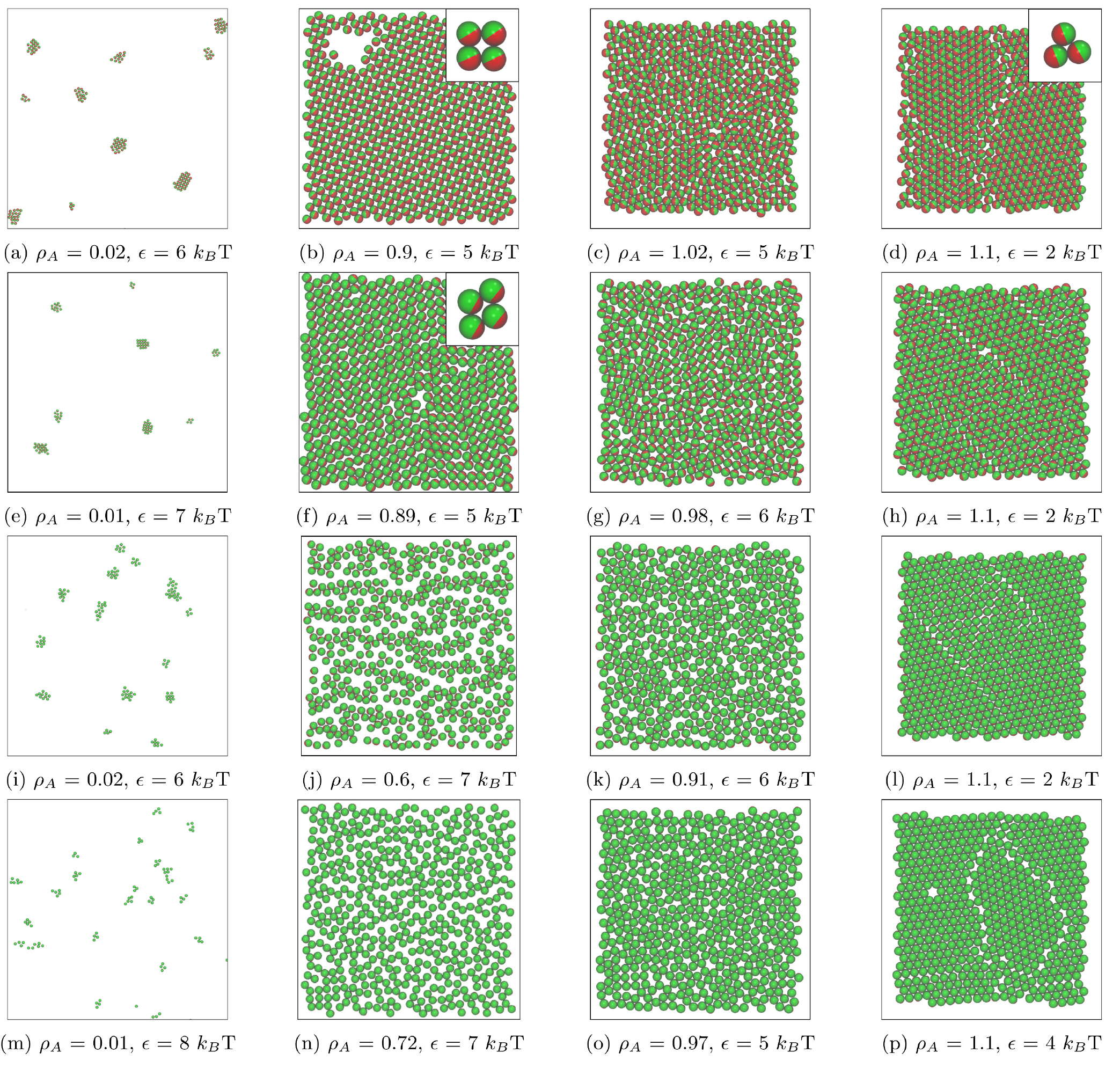}
\captionsetup{justification=raggedright}
\caption{Representative configuration of the system at each phase for different patch coverages. (a)-(d) illustrates the system at $\chi$ = 0.5, (e)-(h) at $\chi$ = 0.33, (i)-(l) at $\chi$ = 0.22, (m)-(p) at $\chi$ = 0.12. Fig.~(a),(e),(i),and (m) depict the cluster phases of the different systems. Fig.~(b) and (f) depict the square and rhombic crystal, (j) and (n) shows the arrested gel phase gel(I)] at $\chi = 0.22$ and gel(II) at $\chi = 0.12$. Fig.~(c),(g),(k) and (o) depict the transition to triangular crystal. Fig.~(d),(h),(l),and (p) depict the triangular crystal formation in the different systems. (color online)}
\label{snapshot}
\end{figure*}

\begin{figure}[t]
\centering
\includegraphics[height=7cm]{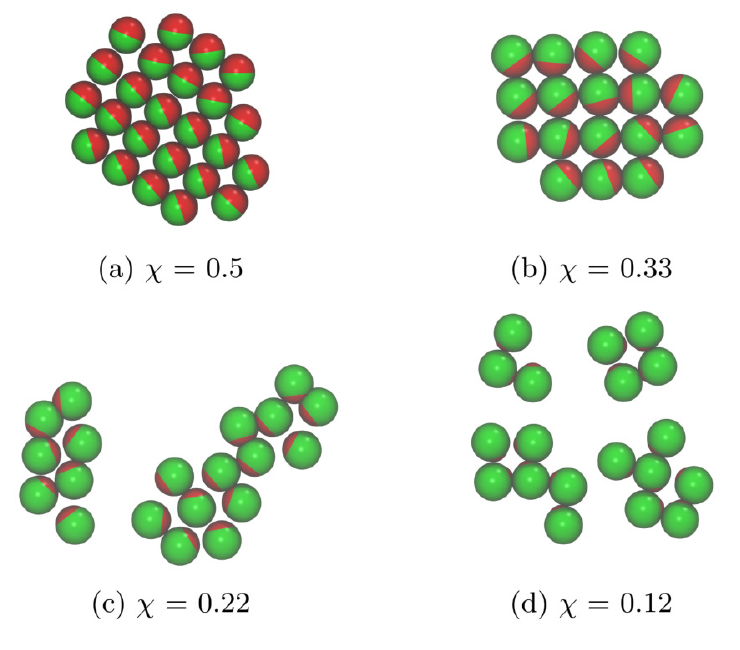}
\captionsetup{justification=raggedright}
\caption{Illustrative images for clusters of different patch coverages at $\rho_A$ = 0.01. (color online)}
\label{cluster}
\end{figure}

Fig.~\ref{statediagram} shows the state diagram for $\chi$ = 0.5, $\chi$ = 0.33, $\chi$ = 0.22 and $\chi$ = 0.12. Particles with patch coverage of 0.5 is equivalent to Janus particles or zwitterionic particles having equal and opposite hemispheres. Here we consider circular discs having two equal halves, one positively charged (or coated with single stranded DNA) half and a negatively charged (or coated with complementary DNA) one.  Fig.~\ref{statediagram}a shows the different phases exhibited by the particles as a function of $\rho_A$ and $\epsilon$. At low values of $\epsilon$ and density, particles are in a gaseous state, while liquid phase is stabilized at moderate densities. The empty region seen indicates a broad coexistence region between the gas and liquid phases which is also reported in the 3D case \cite{dempster2016aggregation}. Kinetically stable finite-sized cluster phase is observed at higher $\epsilon$ values ($\epsilon$ = 5 and 6) and low densities (0.001 $< \rho_A <$ 0.4) as shown in Fig.~\ref{snapshot}a. The particles in the clusters are arranged in a square lattice as shown in Fig.~\ref{cluster}a. The size of the clusters is limited by the availability of monomers and diffusion limitations of the clusters. The average cluster size is found to increase with density as more particles are available for the growth of the cluster. At higher densities (0.6 $< \rho_A <$ 0.9 ) and higher $\epsilon$ values, the particles stabilize a square crystal. This can be seen in Fig.~\ref{snapshot}b. While previous simulations reporting square crystals use 4-patch model \cite{zhang2004self} or IPC with two patches in 2D \cite{bianchi2014tunable}, the present simulation shows IPC with single patch is sufficient to stabilize square crystal. A 3D analogue of a simple cubic crystal is reported to be stabilized by 6-patchy octahedral patchy particles \cite{doye2007controlling}. As the density is increased beyond 0.9 we observe a transition from square crystal to a triangular crystal as shown in Fig.~\ref{snapshot}d. While performing NPT simulations at higher $\epsilon$ values and high pressures, stabilising ordered phases becomes difficult due to poor equilibration. In such cases, we use a pre-generated initial configuration of a square crystal or a triangular crystal and check for the stability of the crystal. The open symbols in the state diagram represent the ordered crystal phases that are rechecked with these stability simulations. We also note a transition region between square and triangular crystal where both crystal domains coexist (Fig.~\ref{snapshot}c).\\

Fig.~\ref{statediagram}b shows the state diagram for particles with 0.33 patch coverage. The phase behavior of particles with a patch coverage of 0.33 is very similar to that of 0.50, as far as gas, liquid, cluster and triangular crystal phases are concerned. For instance, compact 2D crystal clusters showing a uniform distribution of particles are seen at higher $\epsilon$ values and low densities (Fig.~\ref{cluster}b) and the average cluster size increases with density. However, there is one striking difference: the low density crystal takes a rhombic lattice instead of a square lattice.  Representative snapshots of cluster, rhombic crystal and triangular crystal are shown in Fig.~\ref{snapshot}e, \ref{snapshot}f and \ref{snapshot}h respectively. Transition region between rhombic and triangular crystals are also shown in Fig.~\ref{snapshot}g. In literature, rhombic crystal has been shown to form via convective assembly of particles under non-equilibrium conditions in microchannels depending upon the relative ratio of channel width to the particle diameter \cite{kumacheva2003two}. Contrary to this finding, we show here equilibrium rhombic crystal structure being stabilized by tuning patchy interactions of the IPC model. \\

The different phases exhibited by the system of particles with 0.22 patch coverage is shown in Fig.~\ref{statediagram}c as a function of $\rho_A$ and $\epsilon$. Similar to the previous cases, gas and liquid phases are found at the low $\epsilon$ values. For this patch coverage the repulsive interactions due to the non-patch surfaces are found to dominate, restricting the number of bonded neighbors of a particle. As a consequence the clusters found in this case are elongated in shape and self repelling as shown in (Fig.~\ref{snapshot}i, \ref{cluster}c) due to which the average cluster size is a weak function of density. These clusters are kinetically stable and are found for $\epsilon >$ 4, and densities in the range of 0.001 $< \rho_A <$ 0.1. At moderate densities (0.5 $< \rho_A <$ 0.8), these clusters establish connected networks and form a bilayered gel-like arrested state. Fig.~\ref{snapshot}j shows a representative image of the local structure of the gel. Due to the dominance of repulsive interactions, the patchy particles do not stabilise square/rhombic crystals. However, at high densities, triangular crystal is stabilized as shown in Fig.~\ref{snapshot}l. In this range of density, at high $\epsilon$ values the stability of the triangular crystal is checked by performing MCNPT simulations starting with a preformed lattice. \\

At a patch coverage of 0.12, essentially the patch is point-like which leads to increased non-patch vs non-patch repulsions. Though the bare surface of the particle is open to attract more number of particles at low densities, the repulsive interaction overwhelms to limit the number of neighbors at higher densities. This increase in the repulsion also requires the sampling of the system to be done at higher values of $\epsilon$ (upto 8 $k_B$T) to find any stable phase different from liquid. The different phases exhibited by the system as a function of $\rho_A$ and $\epsilon$ are illustrated in Fig.~\ref{statediagram}d. As in the previous cases, gas and liquid structures are observed at the lower $\epsilon$ region of the state diagram. For the higher $\epsilon$ region and at lower densities (0.001 $< \rho_A <$ 0.1), we see clusters consisting of 3, 4, 5 and 6 particles (Fig.~\ref{snapshot}m, \ref{cluster}d) which can be viewed as building blocks of the elongated, self repelling clusters formed at intermediate densities. However, the increase in cluster size is negligible with increase in density due to the limited number of bonded neighbors possible in this condition. At higher density values, these clusters form a highly branched arrested gel phase. It is worth noting that the local gel structure is different between particles with 0.22 and 0.12 patch coverage. Similar to the other three cases, this system also stabilises triangular crystals (Fig.~\ref{snapshot}p) at high densities whose stability at high $\epsilon$ values is studied using the ordered initial configuration mentioned in the previous cases.\\


The discussion above shows that as the patch coverage of the particle is varied, different ordered structures are stabilized. Clusters, square crystal, rhombic crystal, and cross-linked gels of different local structure are stabilized by varying the patch coverage. A triangular crystal is always stabilized at high densities near close-packing limit, irrespective of the patch coverage. The clusters formed at different patch coverages are distinct with respect to their size and shape under identical conditions of density. Further, we discuss below the structural properties of the crystal phases in terms of radial distribution function and $\psi_n$ parameters and the angular properties specifically the polarization and orientational order parameters. 

\subsection{Translational order}

\begin{figure}[h]
\centering
\includegraphics[height=16cm]{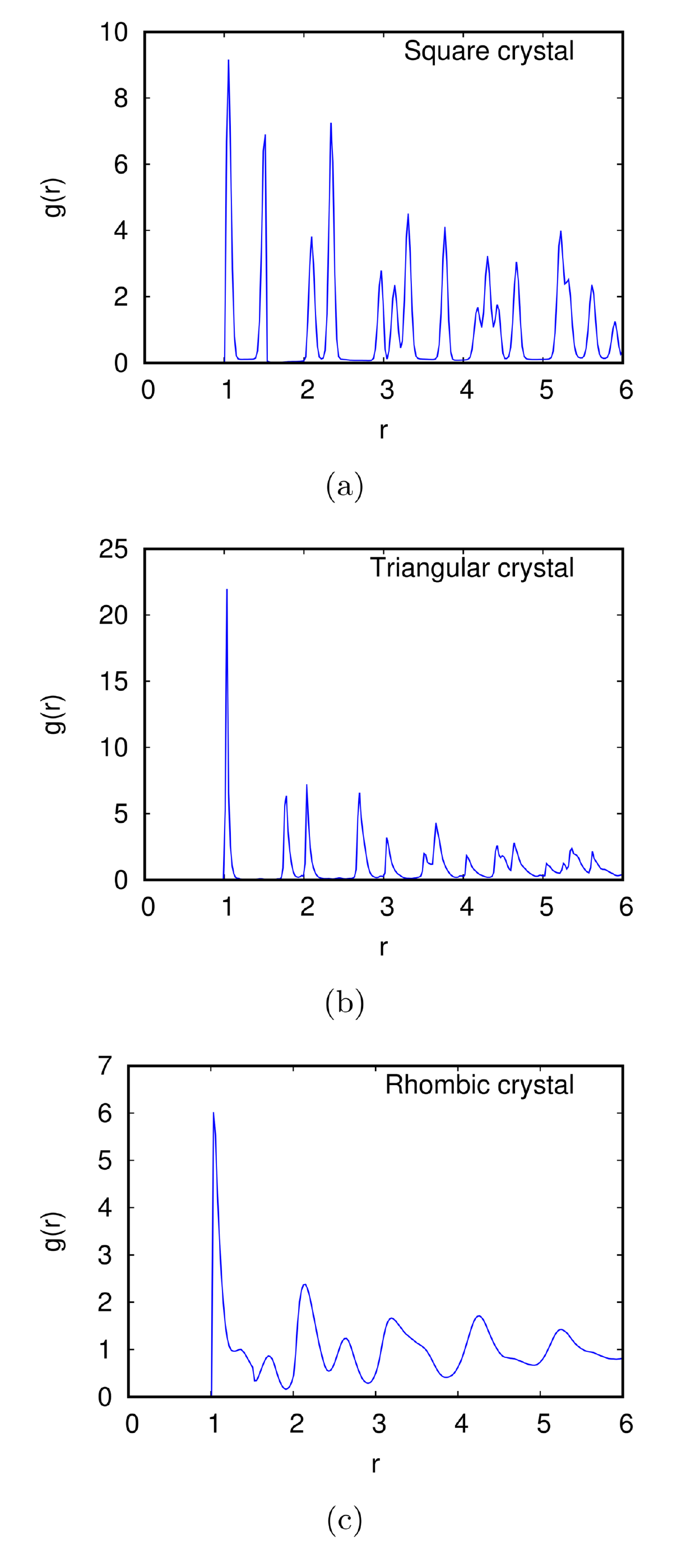}
\captionsetup{justification=raggedright}
\caption{Radial distribution function for the ordered phases formed at different patch coverages as shown in Fig.~\ref{snapshot}.}
\label{rdf}
\end{figure}

The translational order of the various crystal structures can be studied using the radial distribution function (RDF) calculations and the crystal order parameter $\psi_n$. Fig.~\ref{rdf}a shows the RDF for the square crystal observed at the Janus case. The RDF and Fig.~\ref{snapshot}b together demonstrate the high crystallinity of the solid phase. The peaks positions in the RDF follow the trend of $1 : \sqrt2 : 2 : \sqrt5 : 2\sqrt2...$ which is representative of a square crystal. The value of average $\psi_4$ parameter for the observed square crystal is $\approx$ 0.9 which reconfirms the square order of the crystal. Fig.~\ref{rdf}b shows the RDF for the triangular crystal observed at this patch coverage. The change in crystalline nature is confirmed by the change in peak positions in RDF that follows a trend $1 : \sqrt3 : 2 : \sqrt7 : 3...$ which is representative of a triangular crystal. The value of average $\psi_6$ parameter for the observed triangular crystal is $\approx$  0.9. As discussed previously, a reduction in patch coverage to 0.33 stabilises a rhombic crystal. The RDF for the rhombic crystal shown in Fig.~\ref{rdf}c exhibits this change in crystalline order. The value of $\psi_4$ of rhombic crystal is found to be smaller ($\psi_4$ $\approx$ 0.7) than that of square crystal. 
For the system at lower patch coverages ($\chi = 0.22$ \& $0.12$), triangular crystal is stabilized at high densities. At high $\epsilon$ values the stability of the triangular crystal is checked by performing MCNPT simulations starting with a preformed lattice. The triangular crystals stabilised at different patch coverages are found to have similar crystal order at conditions of identical densities.
\subsection{Orientational order}

\begin{figure}[h]
\centering
\includegraphics[height=6cm]{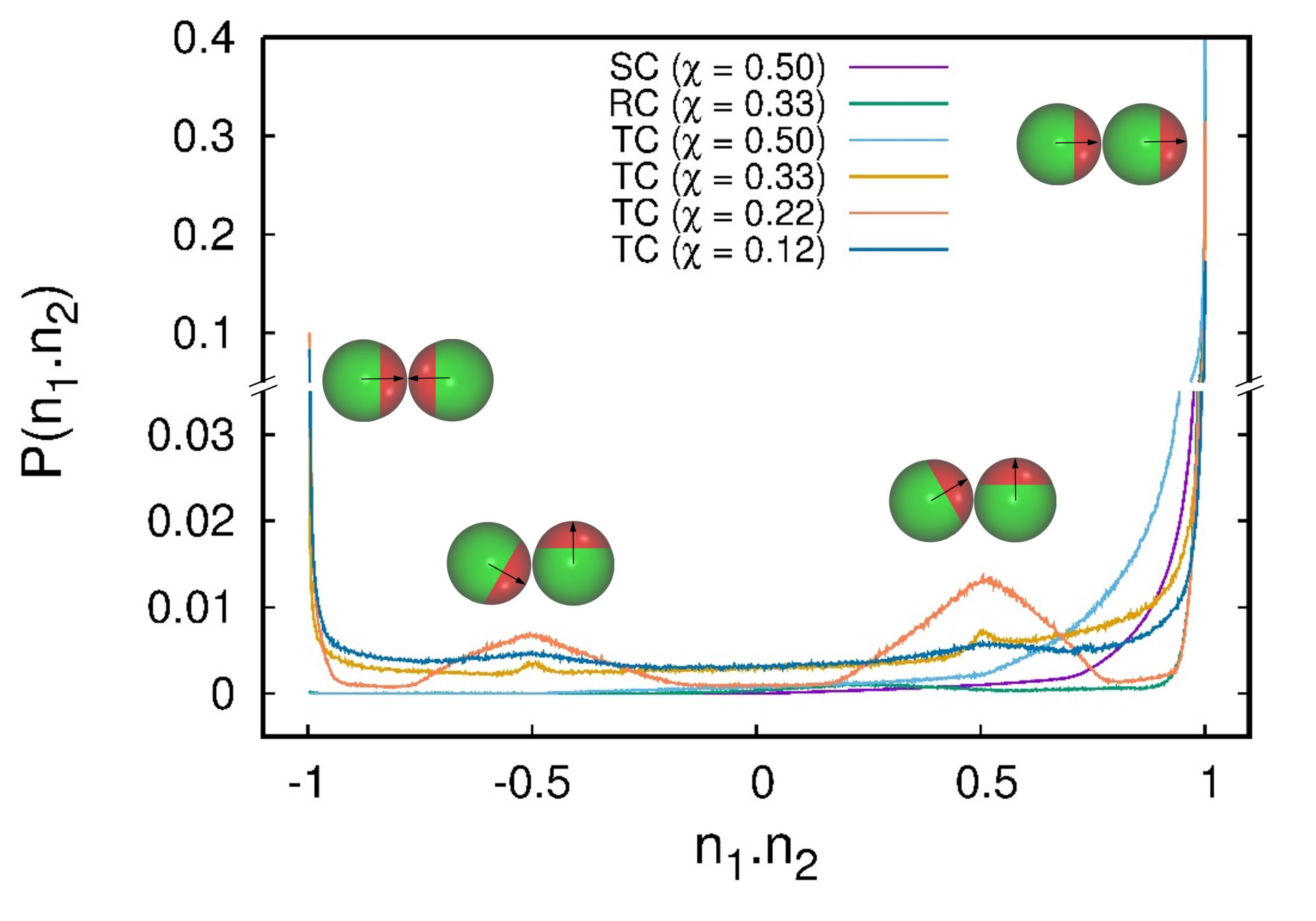}
\captionsetup{justification=raggedright}
\caption{Order parameter calculations showing the relative orientations between particles for triangular crystals and other ordered crystals formed at various patch coverages. Schematic of particle orientations are provided for better understanding. The arrows indicate the unit patch vector. SC, RC and TC stands for square crystal, rhombic crystal and triangular crystal respectively. (color online)}
\label{order}
\end{figure}

The orientational order of the crystal structure is studied using the order parameter and polarization calculations. 
The order parameter P(\textbf{\^{n}}$_1$.\textbf{\^{n}}$_2$) gives the relative orientation between a pair of close packed particles. Let \textbf{\^{n}}$_1$ and \textbf{\^{n}}$_2$ represent the unit patch vectors of particles 1 and 2 within high proximity [$r_{12}\leq  1.05$]. The distribution of the scalar product \textbf{\^{n}}$_1$.\textbf{\^{n}}$_2$, i.e. P(\textbf{\^{n}}$_1$.\textbf{\^{n}}$_2$), is calculated to understand the relative orientations of the particles in different ordered phases \cite{sciortino2009phase,sciortino2010numerical}. The distribution remains flat for an entirely disordered system while it shows distinct peaks for an orientationally ordered phase. The peaks correspond to the most favorable orientations adopted by the pair of bonded particles. Since we observe triangular crystals for all patch coverages at high particle density, first we discuss the particle orientations in the triangular crystal phase. Fig. \ref{order} shows the plot of P(\textbf{\^{n}}$_1$.\textbf{\^{n}}$_2$) for square crystal, rhombic crystal and the triangular crystals obtained for all the patch coverages. For the triangular lattice, we find that the angles between the patch vectors leading to attractive interaction are 0$^{\circ}$, 60$^{\circ}$ and 120$^{\circ}$ corresponding to \textbf{\^{n}}$_1$.\textbf{\^{n}}$_2$ values of 1, 0.5 and -0.5. Fig. \ref{order} shows the corresponding peaks at these angular orientations, confirming self-assembly of particles in a triangular lattice.\\
The figure shows the tendency of the particles to orient themselves in a parallel arrangment corresponding to $\theta=0^{\circ}$ in every system. It is also interesting to note that the particles are also found to orient themselves in an unfavorable antiparallel arrangement in the triangular crystals formed at lower patch coverages. This probability is however much less that the probability to orient in the parallel arrangement. This may be an indication that the triangular crystals are stabilized due to entropy contribution which is independent of the orientation of the particles.
The low density square and rhombic crystals shows an increasing trend towards the patch-non patch orientation from $\theta=90^{\circ}$ to $\theta=0^{\circ}$. This trend is also seen in case of the triangular crystal formed at the Janus case. This may be due to the greater degree of freedom of rotation available for systems with larger patch coverage.

\subsection{Polarization, P}

The measurement of polarization gives the collective orientation of the particles in the system. It can be expressed mathematically as

\begin{equation}
P = \frac{1}{N}\Big|\Big|\sum_{i}\mathbf{\hat n}_{i}\Big|\Big|
\label{eqn3}
\end{equation}

\begin{figure}[t]
\centering
\includegraphics[height=13cm]{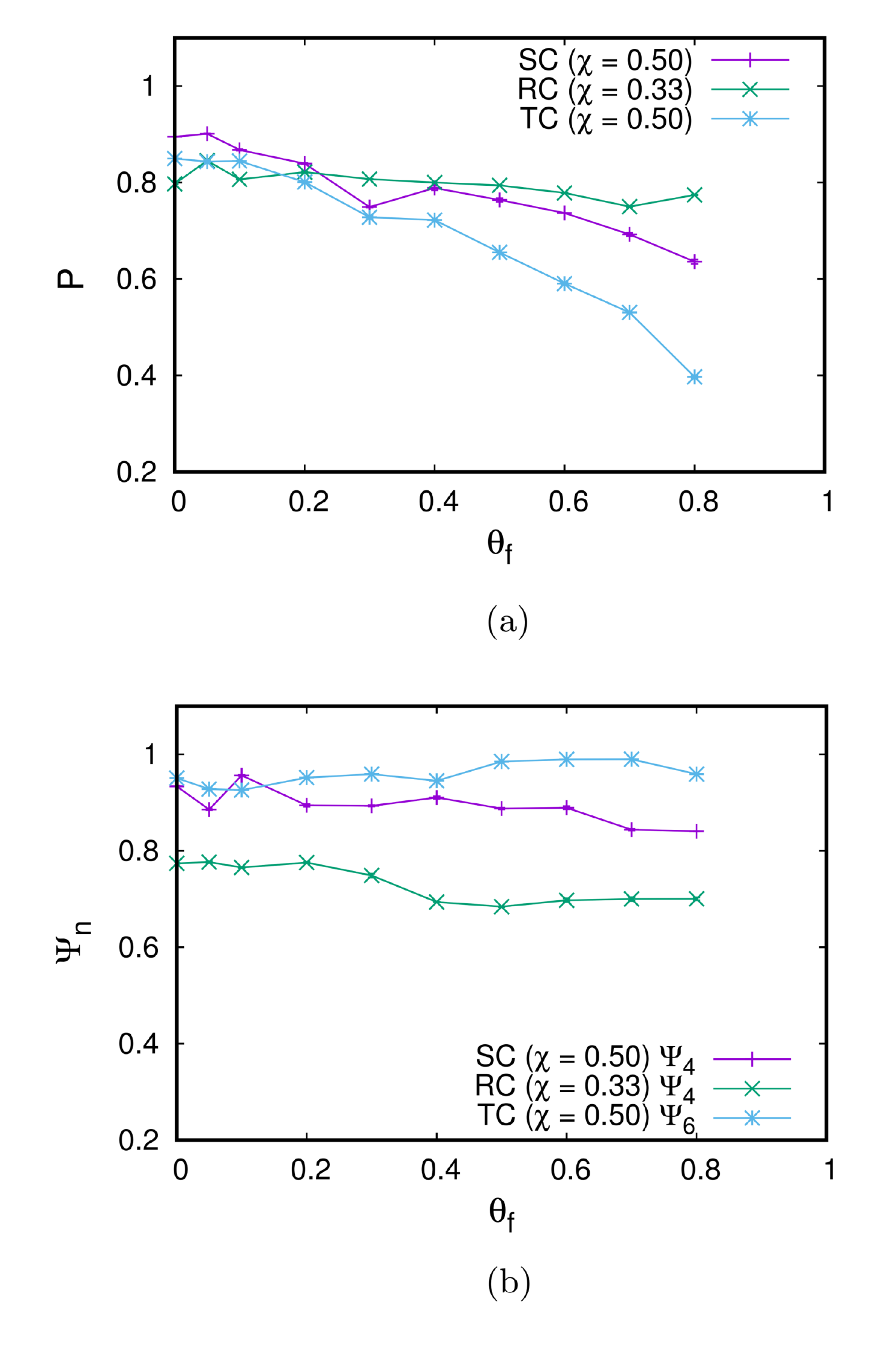}
\captionsetup{justification=raggedright}
\caption{(a)Polarization calculations showing the collective orientations of the particles in the ordered crystals as a function of polydispersity. (b)Variation of $\psi_n$ with polydispersity in patch angle. (color online)}
\label{polydispersity}
\end{figure}

It is already reported in literature that the high degree of polarization of ordered phases are lost with a reduction in patch coverage \cite{dempster2016aggregation}. On a similar line we study the polarization of the ordered crystals found in our system. The study is restricted to the higher patch coverage systems which show a strong polarization at the monodisperse patch condition. We induce a polydispersity in the patch coverage by varying the patch angle $\delta$ of the particles in the crystal phases. This is done by randomly asssigning a value from a uniform distributiom within an interval [$\delta(1-\theta_f), \delta(1+\theta_f$)], where $\theta_f$ is the allowed range of fractional variation in the patch angle. We see a decreasing trend in the value of polarization as the range of polydispersity increases as shown in Fig.~\ref{polydispersity}a. However the translational order of the crystals is not affected considerably. The $\psi_4$ values of the square and rhombic crystals and the $\psi_6$ value of the triangular crystal show good stability with polydispersity as shown in Fig.~\ref{polydispersity}b.

\section{Conclusions}

Using Monte Carlo simulations, the state diagrams of inverse patchy particles with different patch coverages are presented. These patchy particles stabilize regular square crystal, rhombic crystal and close-packed triangular crystals in two-dimensions depending upon the patch coverage, interaction strength and particle density. As these crystals exhibit different lattice structures and lattice parameters, their optical properties such as photonic band gap can be tuned, thus making them useful in optoelectronic devices. The crystals also have pores of different size and shape suited for size and selective membrane-based separation processes. Hence, we see that a single patch model system shows diverse variety of self-assembled structures by merely tuning the patch coverage, particle density and strength of interaction. While the interactions are modelled as orientation dependent square-well and square-shoulder potentials, use of other form of potential is also likely to yield the same result as long as the orientation dependent term is taken into account. The study demonstrates the crucial role of patch coverage on the self-assembled structures stabilized by inverse patchy colloids. These crystal phases may also be realised in experiments if the self-assembly studies are confined to quasi two-dimensions. Whether these crystal types are stable in three dimensions or if more crystal structures can be stablized are yet to be studied. It will be interesting to study the vapor-liquid equilibria in 2D as a function of patch coverage. These issues will be addressed in the future publications.

\acknowledgments

We acknowledge the Virgo High-Performance Computing Environment (HPCE) at IIT-Madras for the computational resources. E. M acknowledges the Department of Science and Technology, India, for the funding through research grant (SR/S3/CE/055/2012).

\bibliographystyle{aipnum4-1}
\bibliography{references}
\end{document}